\documentclass[a4paper,12pt]{article}
\usepackage{amsmath}
\usepackage{pstricks}
\usepackage{pst-node}
\usepackage[ansinew]{inputenc}
\usepackage{amssymb,amsmath}
\usepackage{amsfonts}
\usepackage{epsfig}
\usepackage[english]{babel}
\usepackage{graphics}

\def\p{\partial}

\def\a{\alpha}
\def\b{\beta}
\def\g{\gamma}

\def\o{\omega}
\def\e{\varepsilon}

\font\Sets=msbm10

\def\Integer {\hbox{\Sets Z}}    \def\Real {\hbox{\Sets R}}

   \def\Natural {\hbox{\Sets N}}

 \def\Rational {\hbox{\Sets Q}}

\def\be{\begin{equation}}       \def\ba{\begin{array}}

\def\ee{\end{equation}}         \def\ea{\end{array}}

\def\bea {\begin{eqnarray}}      \def\eea {\end{eqnarray}}

\def\bean{\begin{eqnarray*}}    \def\eean{\end{eqnarray*}}

\def\RA {\ \Rightarrow\ }

\def\<{\langle} \def\({\left(}  \def\>{\rangle} \def\){\right)}

\newtheorem{exi}{Example}

\author{Elena Kartashova\\
RISC, J. Kepler University,\\ Altenbergerstr. 69, 4040 Linz, Austria\\
e-mails: lena@risc.uni-linz.ac.at}

\title{Theory of Laminated Turbulence: Open Questions}

\begin{document}
\date{}
\maketitle


\section{Introduction}
The roots of the theory of nonlinear dispersive waves date back to
hydrodynamics of the 19th century. It was observed, both
experimentally and theoretically, that, under certain circumstances,
the dissipative effects in nonlinear waves become less important
then the dispersive ones. In this way, balance between nonlinearity
and dispersion gives rise to formation of stable patterns (solitons,
cnoidal waves, etc.). Driven by applications in plasma physics,
these phenomena were widely studied, both analytically and
numerically, starting from the middle of the 20th century. The main
mathematical break-through in the theory of nonlinear evolutionary
PDEs was the discovery of the phenomenon of their integrability that
became the starting point of the modern theory of integrable systems
with Korteweg-de Vries equation being the first instance in which
integrability appeared. But most evolutionary PDEs are not
integrable, of course. As a powerful tool for numerical simulations,
the method of kinetic equation has been developed in 1960-th  and
applied to many different types of dispersive evolutionary PDEs. The
wave kinetic equation is approximately equivalent to the initial
nonlinear PDE: it is an averaged equation imposed on a certain set
of correlation functions and it is in fact one limiting case of the
quantum Bose-Einstein equation while the Boltzman kinetic equation
is its other limit. Some statistical assumptions have been used in
order to obtain kinetic equations; the limit of their applicability
then is a very complicated problem which should be solved separately
for each specific equation.\\

The role of the nonlinear dispersive PDEs in the theoretical physics
is so important that the notion of dispersion is used for "physical"
classification of the equations in partial variables. On the other
hand, the only mentioning of the notion "dispersion relation"  in
mathematical literature  we have found in the book of V.I.Arnold
\cite{arnold} who writes about important physical principles and
concepts such as energy, variational principle, the Lagrangian
theory, dispersion relations, the Hamiltonian formalism, etc. which
gave a rise for the development of large areas in mathematics
(theory of Fourier series and integrals, functional analysis,
algebraic geometry and many others). But he also could not find
place for it in the consequent mathematical presentation of the
theory of PDEs and the words "dispersion relation" appear only in
the introduction.
 \\

 In our paper we present wave turbulence theory as a base of the "physical" classification of PDEs,
 trying to avoid as much as possible specific physical jargon and give a "pure" mathematician a
 possibility to
follow its general ideas and results. We show that the main
mathematical object of the wave turbulence theory  is an algebraic
system of equations  called resonant manifolds. We also present here
the  model of the laminated wave turbulence that includes classical
statistical results on the turbulence as well as the results on the
discrete wave systems. It is shown that discrete characteristics of
the wave systems can be described in terms of integer points on the
rational manifolds which is the main novelty of the theory of
laminated turbulence. Some applications of this theory for
explanation of important physical effects are given. A few open
mathematical and numerical problems are formulated at the end. Our
purpose is attract pure mathematicians to work on this subject.

\section{General Notions}

For the complicity of presentation we began this section with a very
brief sketch of the traditional mathematical approach to the
classification of PDEs.

\subsection{Mathematical Classification}
Well-known mathematical classification of PDEs is based on {\bf the
form of equations} and can be briefly presented as follows. For a
bivariate PDE of the second order
$$ a \psi_{xx} +b \psi_{xy} +
c  \psi_{yy} =F(x,y,\psi, \psi_x, \psi_y)$$ its characteristic
equation is written as $$\frac{\boldsymbol{d} x}{\boldsymbol{d} y}=
\frac{b}{2a} \pm \frac{1}{2a}\sqrt {b^{2}-4ac}$$ and three types of
PDEs are defined:
\begin{itemize}
\item{} $b^2<4ac$, elliptic PDE:  $ \psi_{xx}+\psi_{yy} = 0$
\item{} $b^2 >4ac$,
hyperbolic PDE: $\psi_{xx}-\psi_{yy} - x\psi_x = 0$
\item{} $b^2=4ac$,
parabolic PDE:  $\psi_{xx}-2xy\psi_y - \psi = 0$
\end{itemize}
Each type of PDE demands then special type of initial/boundary
conditions for the problem to be well-posed.
 "Bad"  example of Tricomi equation $y\psi_{xx}+\psi_{yy} = 0$
 shows immediately incompleteness  of this classification even for second order
 PDEs because
  a
 PDE can change its type depending, for instance, on the initial
 conditions. This classification can be generalized to PDEs of more
 variables but not to PDEs of higher order.

\subsection{Physical Classification}Physical classification of PDEs
is based on {\bf the form of solution} and is almost not known to
pure mathematicians. In this case, a PDE is regarded in the very
general form, without any restrictions on the number of variables or
the order of equation. On the other hand, the necessary preliminary
step in this classification is the division of all the variables
into two groups - time- and space-like variables. This division
originated from the special relativity theory where time and
three-dimensional space are treated together as a single
four-dimensional  Minkowski space. In Minkowski space a metrics
allowing to compute an interval $s$ along a curve between two events
is defined analogously to distance in Euclidean space:
$$
\boldsymbol{d} s^2 = \boldsymbol{d} x^2 + \boldsymbol{d} y^2 +
\boldsymbol{d}z^2  - c^2 \boldsymbol{d} t^2$$
 where $c$ is speed
of light, $x,y,z$ and $t$ denote respectively space and time
variables. Notice that though in mathematical classification all
variables are treated equally, obviously its results can be used in
any applications only after similar division of variables have been
done.\\

Suppose now
 that linear PDE with constant coefficients
 has a wave-like solution
$$
\psi (x,t)= A \exp {i[ kx - \o t]} \quad \mbox{or} \quad \psi (x,t)=
A \sin ( kx - \o t)
$$
with amplitude $A$, wave-number $k$ and wave frequency $\o$. Then
the substitution of $\quad \p_t =-i  \o , \quad \p_x = i k  \quad$
transforms LPDE  into a {\bf polynomial} on $\o$ and $k$, for
instance:
$$
\psi_t+\a\psi_x+\b\psi_{xxx}=0 \quad \RA  \o ( k)=\a k - \b k^3,
$$
$$
\psi_{tt} + \a^2 \phi_{xxxx}=0 \quad \RA  \o^2 ( k)=  \a^2  k ^4,
$$
$$
\psi_{tttt}-\a^2\psi_{xx}+\b^2\psi=0 \quad \RA  \o^4 ( k)=
\a^2k^2+\b^2
$$
where $\a$ and $\b$ are constants.

\paragraph{Definition  } Real-valued function $\quad \o=\o(k): \ \ \ {\rm d}^2 \o /{\rm
d} k^2  \neq 0 \quad $ is called {\it dispersion relation} or {\it
dispersion function}. A linear PDE with wave-like solutions are
called evolutionary dispersive LPDE.  A nonlinear PDE with
dispersive linear part are called
 evolutionary dispersive NPDE.\\

This way all PDEs are divided into two classes - dispersive and
non-dispersive \cite{whith1}. This classification is not
complementary to a standard mathematical one. For instance, though
hyperbolic PDEs normally do not have dispersive wave solutions, the
hyperbolic equation $\psi_{tt}- \a^2  \psi_{xx} + \b^2 \psi =0 $ has
them. Given dispersion relation allows to re-construct corresponding
linear PDE. All definitions above could be easily reformulated for a
case of more space variables, namely $x_1, x_2,...,x_n$.
 Linear part of the initial PDE takes then form
$$
P(\frac{\p}{\p t},\frac{\p}{\p x_1},...,\frac{\p}{\p x_n})
$$
and correspondingly dispersion relation can be computed from
$$
P(-i\o, ik_1,...,ik_n)=0$$ with the polynomial $P$.
 In this case we will have not a
wave number $k$ but a {\it wave vector} $\vec k = (k_1,...,k_n)$ and
the condition of non-zero second derivative of the dispersion
function takes a matrix form:
$$
\arrowvert \frac {\partial^2\omega}{\partial k_i \partial k_j}
\arrowvert \neq 0.
$$

\subsection{Perturbation technique}
Perturbation or asymptotic methods (see, for instance, \cite{ho})
are much in use in
 physics and
are dealing with equations having some small parameter $\e
 >0$. To understand the results  presentated in the next Section one needs to have some clear idea
 about  the perturbation technique and this is the reason why we
give here a simple algebraic example of its application. The main
idea of  a perturbation method is very straightforward - an unknown
solution, depending on $\e $, is written out in a form of infinite
series on different powers of $\e$ and coefficients in front of any
power of $\e$ are computed consequently. Let us take an algebraic
equation \be \label{pert}x^2 - (3-2\e)x + 2 + \e =0 \ee and try to
find its asymptotic solutions. If $\e = 0$ we get \be
\label{unpert}x^2 - 3x + 2 =0 \ee with roots $x=1$ and $x=2$.
Eq.(\ref{pert}) is called {\it perturbed} and Eq.(\ref{unpert}) -
{\it unperturbed.} Natural suggestion is that the solutions of
perturbed equation differ only a little bit from the solutions of
unperturbed one. Let us look for solutions of Eq.(\ref{pert}) in the
form
$$
x = x_0 + \e x_1 + \e ^2 x_2 + ... $$ where $x_0$ is a solution of
Eq.(\ref{unpert}), i.e.  $x_0=1$ or $x_0=2$. Substituting
 this infinite series into Eq.(\ref{pert}), collecting
 all the terms with the same
degree of $\e$  and consequent equaling to zero all coefficients in
front of different
 powers of $\e$ leads to an algebraic system of equations
\bea \label{eps}
\begin{cases}
\e^0: x_0^2 - 3x_0 + 2=0, \\
\e^1: 2x_0 x_1 - 3x_1 - 2 x_0 +1=0,  \\
\e^2: 2x_0x_2 + x_1^2 -3x_2 - 2x_1=0,  \\
...
\end{cases}
\eea
 with solutions
$$
x_0 = 1, \quad x_1 = -1, \quad x_2 = 3, ...  \quad \mbox{and} \quad
x = 1 - \e + 3 \e ^2 + ...;
$$
$$
x_0 = 2, \quad x_1 = 3, \quad x_2 = -3, ... \quad \mbox{and} \quad x
= 2 + 3\e - 3 \e ^2 + ...$$ Notice that
 exact solutions of Eq.(\ref{pert}) are
$$
x = \frac {1}{2} [3 + 2 \e \pm \sqrt {1 + 8 \e + 4 \e ^2}]
$$
and  the use of binomial representation for the expression under the
square root
$$
(1 + 8 \e + 4 \e ^2)^ \frac{1}{2} = 1 + (8 \e + 4 \e ^2) + \frac
{\frac{1}{2} (\frac {-1}{2})}{2!} (8 \e + 4 \e ^2)^2 + ... =  1 +
4\e - 6 \e ^2 + ...
$$ gives finally
$$
x = \frac{1}{2} (3 + 2 \e  - 1 - 4 \e  + 6 \e  ^2 + ...)  = 1 - \e +
3 \e  ^2 + ...$$ and
$$ x = \frac{1}{2} ( 3 + 2 \e  + 1 + 4\e  - 6 \e  ^2 + ...) =
 2 + 3\e  - 3 \e  ^2 + ...$$ as before.
This example was chosen because the exact solution in this case is
known and can be compared to the asymptotic one. The same approach
is used for partial differential equations, also in the cases when
exact solutions are not known. The only difference would be more
elaborated  computations resulting in some system of ordinary
differential equations instead of Eqs.(\ref{eps}) (see next
Section).
\section{Wave Turbulence Theory}

\subsection{Wave Resonances}

Now we are going to introduce the notion of the wave resonance which
is the mile-stone for the whole theory of evolutionary dispersive
NPDE and therefore for the wave turbulence theory. Let us consider
first a linear oscillator driven by a small force
$$
x_{tt} + p^2 x = \e e^{i \Omega t}.$$
 Here $p$ is eigenfrequency
of the system, $\Omega$ is frequency of the driving force and $\e
>0$ is a small parameter. Deviation of this system from
equilibrium is small (of order $\e$), if there is no resonance
between the frequency of the driving force $\e e^{i \Omega t}$ and
an eigenfrequency of the system. If these frequencies coincide then
the amplitude of oscillator grows linearly with the time and this
situation is called {\it resonance} in physics. Mathematically it
means
{\it existence of unbounded solutions}.\\

Let us now regard a (weakly) nonlinear PDE of the form 
\be \label{Non_eps}L(\psi)=\e N(\psi)\ee
 where $L$ is an arbitrary linear
dispersive operator and $N$ is an arbitrary nonlinear operator. Any
two  solutions of $L(\psi)=0$ can be written out as
$$
A_1 \exp{i[\vec k_1 \vec x -\omega (\vec k_1) t]} \quad \mbox{and}
\quad A_2 \exp{i[\vec k_2 \vec x -\omega (\vec k_2) t]}$$
 with constant
amplitudes $A_1, A_2$. Intuitively natural expectation is that
solutions of weakly nonlinear PDE will have the same form as linear
waves but perhaps with amplitudes depending on time. Taking into
account that nonlinearity is small, each amplitude is regarded as
{\it a slow-varying function of time}, that is $A_j = A_j(t/\e)$.
Standard notation is $A_j=A_j(T)$ where $T=t/\e$ is called {\it slow
time}. Since wave energy is by definition proportional to
amplitude's square $A_j^2$ it means that in case of nonlinear PDE
{\it waves exchange their energy}. This effect is also described as
"waves are interacting with each other" or "there exists energy
transfer
through the wave spectrum" or similar. \\

Unlike linear waves for which their linear combination was also
solution of $L(\psi)=0$, it is not the case for nonlinear waves.
Indeed, substitution of two linear waves into the operator $\epsilon
N(\psi)$ generates terms of the form $ \exp{i[(\vec k_1 +\vec k_2)
\vec x -[\o (\vec k_1) + \omega (\vec k_2)] t} $
 which play the role of a small driving
force for the linear wave system  similar to the case of linear
oscillator above. This driving force gives a small effect on a wave
system till resonance occurs, i.e. till the wave number and the wave
frequency of the driving force does not coincide with some wave
number and some frequency of eigenfunction:
 \bea \label{3res}
 \begin{cases}
\o_1 + \o_2= \o_3,\\
 \vec k_1
+ \vec k_2 = \vec k_3
\end{cases}
\eea 
where notation $\o_i=\omega (\vec k_i)$ is used. This system
describes so-called {\it resonance conditions} or {\it
resonance manifold}.\\

The perturbation technique described above produces the equations
for the amplitudes of resonantly interacting waves $A_j=A_j(T)$. Let
us demonstrate it taking as example barotropic vorticity equation
(BVE) on a sphere
\begin{equation}\label{BVE}
\frac{\partial \triangle \psi}{\partial t} + 2 \frac{\partial
\psi}{\partial \lambda} + \e  J(\psi,\triangle  \psi) =0
\end{equation}
where
$$
\triangle \psi = \frac{\partial^2 \psi}{\partial \phi^2}+ \frac
{1}{\cos^2 \phi} \frac {\partial^2 \psi}{\partial \lambda^2} - \tan
\phi \frac{\partial \psi}{\partial \phi} \quad \mbox{and} \quad
J(a,b)= \frac{1}{cos \phi}(\frac{\partial a}{\partial \lambda}
\frac{\partial b}{\partial \phi} - \frac{\partial a}{\partial \phi}
\frac{\partial b}{\partial \lambda}).$$ The linear part of spherical
BVE has wave solutions in the form
\begin{equation} \label{math_shock}
A P_n^m (\sin \phi) \exp{i[m \lambda +\frac{2m}{n(n+1)}
t]},\nonumber
\end{equation}
where $A$ is constant wave amplitude,  $\ \omega = -2m /[n(n+1)]\ $
and $ \ P_n^m (x) \ $ is the
associated Legendre function of degree $n$ and order $m$.\\

One of the reasons to choose this equation as an example is
following. Till now a linear wave was supposed to have much more
simple form, namely, $A \exp{i[\vec k \vec x -\omega (\vec k) t]}$
without any additional factor of a functional form. For a physicist
it is intuitively clear that if the factor is some {\it oscillatory
function of only space variables} then we will still have a wave of
a sort "but it would be difficult to include it in an overall
definition. We seem to be left at present with the looser idea that
whenever oscillations in space are coupled with oscillation in time
through a dispersion relation, we expect the typical effects of
dispersive waves" \cite{whith1}.
By the way, most physically important dispersive equations have the waves of  this form.\\

Now let us keep in mind that a wave is something more complicated
then just a $\ \sin \ $ but still smooth and periodic, and  let us
look where perturbation method will lead us. An approximate solution
has a form
\begin{equation}
\psi = \psi_0(\lambda, \phi, t, T) + \e
 \psi_1(\lambda,
\phi, t, T)+ \e ^2 \psi_2(\lambda, \phi, t, T) + ...\nonumber
\end{equation}
where $T=t/\e $ is the slow time and the zero approximation $\psi_0$
is given as a sum of three linear waves:
\begin{equation}
\psi_0(\lambda, \phi, t, T) = \sum_{k=1}^3 A_k(T)P^{(k)} \exp(i
\theta_k)
\end{equation}
with notations $P^{(k)} = P_{n_k}^{m_k}$ and $\theta _k = m_k
\lambda - \omega_k t$. Then
\begin{eqnarray}
\begin{cases}
\e
 ^0: \quad \partial \triangle \psi_0/\partial t + 2 \partial \psi_0/\partial \lambda=0,  \\
\e
 ^1: \quad \partial \triangle \psi_1/\partial t + 2 \partial \psi_1/\partial \lambda= -J(\psi_0,\triangle \psi_0)- \partial \triangle \psi_0/\partial T, \\
\e
 ^2: .....
 \end{cases}
\end{eqnarray}
and
\begin{eqnarray}
\frac{\partial \triangle \psi_0}{\partial \lambda} = i\sum_{k=1}^3 P^{(k)}m_k N_k[A_k \exp(i\theta_k)-A_k^*\exp(-i\theta_k)];\nonumber \\
\frac{\partial \triangle \psi_0}{\partial \phi} = -\sum_{k=1}^3 N_k\frac {d}{d \phi}P^{(k)}\cos \phi[A_k \exp(i\theta_k)+A_k^*\exp(-i\theta_k)];\nonumber \\
J(\psi_0, \triangle \psi_0)= \nonumber \\
-i \sum_{j,k=1}^3 N_k m_j P^{(j)}\frac {d}{d \phi} P^{(k)}[ A_j
\exp(i\theta_j)-A_j^*\exp(-i\theta_j)]
[A_k \exp(i\theta_k)+A_k^*\exp(-i\theta_k)]+\nonumber \\
+i \sum_{j,k=1}^3 N_k m_k P^{(k)}\frac {d}{d \phi} P^{(j)}[ A_j
\exp(i\theta_j)+A_j^*\exp(-i\theta_j)]
[A_k \exp(i\theta_k)-A_k^*\exp(-i\theta_k)];\nonumber \\
\frac{\partial \triangle \psi_0}{\partial T} = \sum_{k=1}^3
P^{(k)}N_k[\frac {d A_k }{d T}\exp(i\theta_k)+\frac{d A_k^*}{d
T}\exp(-i\theta_k)]. \nonumber
\end{eqnarray}
leads to the condition of unbounded growth of the left hand  in the
form
$$
J(\psi_0, \triangle \psi_0) = \frac{\partial \triangle
\psi_0}{\partial T} $$ with resonance conditions $ \theta_j
+\theta_k =\theta_i \quad \forall \ j,k,i=1,2,3. $ Let us fix some
specific resonance condition, say, $\ \theta_1 +\theta_2 =\theta_3,
\ $ then
\begin{eqnarray}
\frac{\partial \triangle \psi_0}{\partial T} \cong
-N_3 P^{(3)} [\frac{d A_3}{d T} \exp (i \theta_3)+ \frac{d A_3^*}{d T} \exp (-i \theta_3)],\nonumber \\
J(\psi_0, \triangle \psi_0) \cong  -i(N_1-N_2)
(m_2P^{(2)}\frac{d}{d \phi}P^{(1)})-m_1P^{(1)} \frac{d}{d \phi}P^{(2)} \cdot \nonumber \\
{A_1A_2 \exp[i(\theta_1 + \theta_2)]- A_1^*A_2^* \exp[-i(\theta_1 +
\theta_2)]}, \nonumber
\end{eqnarray}
where notation $\cong$ means that only those terms are written out
which can generate chosen resonance. Let us substitute these
expressions into the coefficient by $\e^1$, i.e. into the equation
\begin{equation}
\frac{\partial \triangle \psi_1}{\partial t} + 2 \frac{\partial
\psi_1}{\partial \lambda}= -J(\psi_0,\triangle \psi_0)-
\frac{\partial \triangle \psi_0}{\partial T},\nonumber
\end{equation}
multiply both parts of it by
\begin{equation}
P^{(3)} \sin \phi [A_3 \exp(i\theta_3) + A_3^* \exp(-i\theta_3)]
\nonumber
\end{equation}
and integrate all over the sphere with $t \rightarrow \infty$. As a
result following equation can be obtained: 
\be N_3 \frac{d A_3}{d T}= 2iZ(N_2-N_1)A_1A_2, \nonumber \ee 
 where
\begin{equation}
Z= \int_{-\pi/2}^{\pi/2}[m_2P^{(2)}\frac{d}{d \phi}P^{(1)}-
m_1P^{(1)}\frac{d}{d \phi}P^{(2)}] \frac{d}{d \phi}P^{(3)} d \phi.
\nonumber
\end{equation}
The same procedure obviously provides the analogous equations for
$A_2$ and $A_3$ while fixing corresponding resonance conditions:
\begin{eqnarray}
N_1 \frac{d A_1}{d T}= -2iZ(N_2-N_3)A_3A_2^*, \nonumber\\
N_2 \frac{d A_2}{d T}= -2iZ(N_3-N_1)A_1^*A_3, \nonumber\\
\end{eqnarray}

In general, the simplest system of equations on the amplitudes of
three resonantly interacting waves is often regarded in the form
 \bea \label{complexA}
\begin{cases}
\dot{A}_1=  \a_1 A_3A_2^*, \\
\dot{A}_2=  \a_2 A_1^*A_3, \\
\dot{A}_3=  \a_3 A_1A_2 \,,
\end{cases}
 \eea
and is refereed to as a 3-wave system (keeping in mind that
analogous system has to be written out for $\ A_i^* \ $).
Coefficients $\a_i$ depend on the initial NPDE. Similar system of
equations can be obtained for 4-wave interactions, with the products
of three different amplitudes on the right hand, and so on.

\subsection{Zakharov-Kolmogorov energy spectra}
The idea that a dispersive wave system contains many resonances and
wave interactions are stochastic led to the statistical theory of
wave turbulence. This theory is well developed \cite{lvov} and
 widely used in oceanology and plasma physics  describing a lot
 of turbulent transport phenomena.
 Avoiding the language of Hamiltonian systems, correlators of a wave
 field, etc., one can formulate its main results in the following way. Any nonlinearity
in Eq.(\ref{Non_eps}) can be written out as
\be\label{sigma} \Sigma_{i} \frac{V_{(12..i)} \delta (\vec k_1 +\vec
k_2 + ... +\vec k_{i})} {\delta (\o_1+ \o_2+... + \o_i) } A_1A_2
\cdots A_i \ee
where $\delta$ is Dirac delta-function and $V_{(12..i)}$ is a vertex
coefficient.
 This
presentation, together with some additional statistical suggestions,
is used then to construct a wave kinetic equation, with
corresponding vertex coefficients and delta-functions in the
under-integral expression, of the form
$$
\dot{A}_1=\int |V_{(123)}|^2
\delta(\o_1-\o_2-\o_3)\delta(\vec{k}_1-\vec{k}_2-\vec{k}_3)(A_2A_3-A_1A_2-A_1A_3)
{\bf d}\vec{k}_2 {\bf d}\vec{k}_3
$$
for 3-waves interactions, and similar for $i$-waves interactions.
One of the most important discoveries in the statistical wave
turbulence theory are stationary exact solutions of the kinetic
equations first found in \cite{fil}. These solutions are now called
Zakharov-Kolmogorov (ZK) energy spectra and they describe energy
cascade in the wave field. In other words, energy of the wave with
wave vector $\vec{k}$ is proportional to $k^{\a}$ with $\a<0$ and
magnitude of $\a $ depends on the specific of the wave system.
Discovery of ZK spectra played  tremendous role in the wave
turbulence theory and till the works of last decade \cite{discrete}
it was not realized that some turbulent effects are not due to the
statistical properties of a wave field and are not described by
kinetic equations or ZK energy spectra.

\subsection{Small Divisors Problem}

In order to use presentation (\ref{sigma}) one has to check whether
so defined nonlinearity is finite. This problem is known as {\it the
small divisors problem} and its solution depends on whether wave
vectors
have  real or integer coordinates.\\

Wave systems with {\it continuous spectra} were studied by
Kolmogorov, Arnold and Moser \cite{KAM} and main results of
KAM-theory can be briefly formulated as follows. If dispersion
function $\o$ is defined on real-valued wave vectors and the ratio
$\a_{ij}=\o_i/\o_j$ {\it is not a rational number} for any two wave
vectors $\vec{k}_i$ and $\vec{k}_j$, then

\begin{itemize}
\item{}(1C) Wave system is decomposed into disjoint invariant sets (KAM
tori) carrying quasi-periodic motions;

\item{} (2C) If the size of the wave system tends to infinity, (1C) does not
contradict ergodicity, random phase approximation can be assumed,
kinetic equations and ZK energy spectra describe the wave system
properly;

\item{} (3C) Union of invariant tori has positive Liouville measure and
$\Rational$ has measure 0, there exclusion of the waves with
rational ratio of their dispersions is supposed to be not very
important.
\end{itemize}

Wave systems with {\it discrete spectra}  demonstrate
\cite{discrete} substantially different energetic behavior:
\begin{itemize}
\item{}(1D)
Wave system is decomposed into disjoint discrete classes carrying
periodic motions or empty; for any two waves with wave vectors
$\vec{k}_i$ and $\vec{k}_j$ belonging to the same class, the ratio
$\a_{ij}=\o_i/\o_j$ {\it is a rational number};

\item{}(2D) Energetic behavior of the wave system does not depend on
its size, is not stochastic and is described by a few isolated
periodic processes governed by Sys.(\ref{complexA});

\item{}(3D) In many wave systems (for instance, in the systems with periodic or
zero boundary conditions)  {\it KAM-tori do not exist} and {\it
discrete classes play the major role in the energy transfer}.

\end{itemize}

\subsection{Laminated Wave Turbulence}
The results formulated in the previous section gave  rise to the
model of laminated wave turbulence \cite{JETP} which includes two
co-existing layers of turbulence in a wave system
 - continuous and  discrete layers, each demonstrating specific energetic
 behavior. In other words, KAM-theory describes the wave systems
 leaving some "holes" in the wave spectra which are "full-filled" in
 the theory of laminated turbulence.\\

 Continuous layer, with its kinetic equation, energy cascades, ZK spectra, etc. is well-studied
 while  the existence of the discrete layer was realized quite recently. In order to understand
 which manifestations of the discrete layer are to be expected in
 numerical or laboratory experiments, let us regard an example with dispersion function
 $\ \o=1/\sqrt{m^2+n^2}\ $.
First of all, it is important to realize: the fact that the ratio $\
\a_{ij}=\o_i/\o_j\ $ is a rational number {\it does not imply} that
dispersion function $\o$ is a rational function. Indeed, for $\
\o=1/\sqrt{m^2+n^2} \ $ we have
$$
\o: \Integer \times \Integer \rightarrow \Real
$$
and for wave vectors  $\  \vec{k}_1=(2,1)\  $ and $\
\vec{k}_2=(9,18), \ $ the ratio $\ \o_1/\o_2=1/3 \ $ is rational
number though $\ \o \ $ is irrational function of integer variables.
Decomposition of all discrete waves into disjoint discrete classes
$\ Cl_q  \ $ in this case has the form $$\  \{\vec{k}_i=(m_i,n_i)\}
\in Cl_q  \ \mbox{ if  }  \ |\vec{k}_i|=\g_i \sqrt{q}, \ \forall \
i=1,2,...\ $$ where $\g_i$ is some integer and $q$ is {\it the same
square-free integer}  for all wave vectors of the class $Cl_q$. The
equation $\ \o_1+\o_2=\o_3 \ $ has solutions only if $$\exists
\tilde{q}: \ \vec{k}_1,\ \vec{k}_2, \ \vec{k}_3 \in
Cl_{\tilde{q}}.$$ This is {\it necessary condition}, not sufficient.
The use of this necessary condition allows to cut back substantially
computation time needed to find solution of irrational equations in
integers. Namely, one has to construct classes first and afterwards
look for the solutions among the waves belonging to the same class.
In this way, instead of solving {\it the irrational equation on 6
variables} \be \label{6_irra}
1/\sqrt{m_1^2+n_1^2}+1/\sqrt{m_2^2+n_2^2}=1/\sqrt{m_3^2+n_3^2} \ee
it is enough to solve {\it the rational equation on 3 variables} \be
\label{3_ra} 1/\g_1+1/\g_2=1/\g_3. \ee Some classes can be empty,
for instance $Cl_2$ and $Cl_3$ in our example. Obviously, each
non-empty class has infinite number of elements due to the existence
of proportional vectors so that $\ \vec{k}_1=(2,1)\ $ with the norm
$\ |\vec{k}_1|=\sqrt{5} \ $  and its proportional $\
\vec{k}_2=(9,18) \ $ with the norm $\ |\vec{k}_2|=9 \sqrt{5} \ $
belong to the same class $Cl_5$. On the other hand, not all the
elements of a class are parts of some solution which means that {\it
not all waves take part in resonant
interactions}.\\

Let us come back to physical interpretation of these results.
Resonantly interacting  waves will change their amplitudes according
to Sys.(\ref{complexA}). In this case the role of ZK spectra $\
k^{\a}, \ \a<0, \ $ is played by the interaction coefficient $\ Z
\sim k^{\a}, \ \a>0 \ $  (see Fig.1). Non-interacting waves will
have constant amplitudes (they are not shown in Fig.1). In the next
Section we demonstrate some examples of different wave systems whose
behavior is explained by the theory of laminated turbulence.

\begin{figure}
\includegraphics[width=15cm,height=8cm]{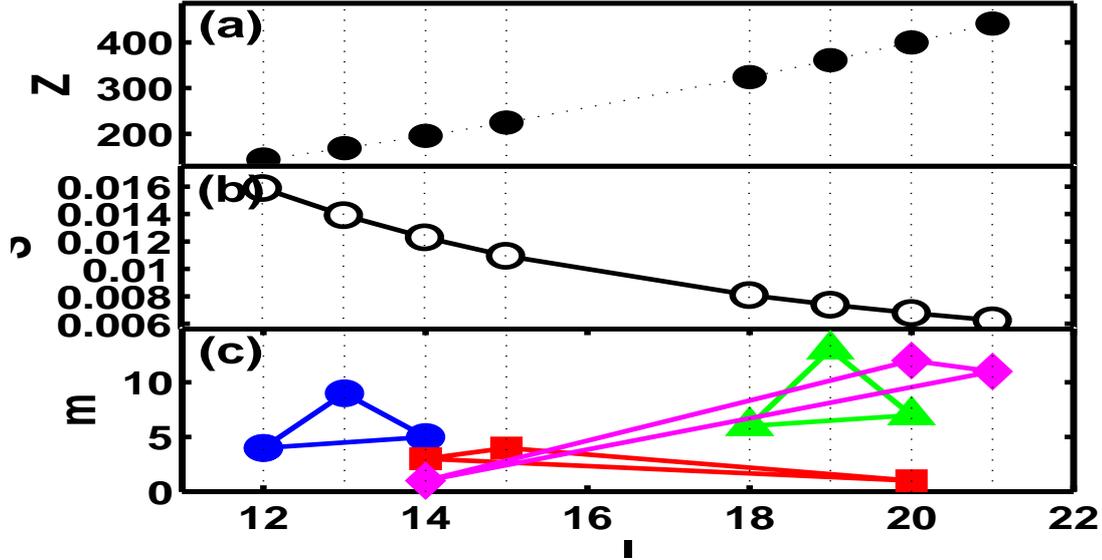}
\caption{Two layers of turbulence are shown symbolically. Low panel:
2D-domain in spectral space, nodes of the integer lattice are
connected with the lines which correspond to 3-wave resonant
interactions  of the discrete layer. Middle panel: ZK energy
spectrum  $\ k^{-3/2} \ $ with "the holes" in the nodes of the
integer lattice. Upper panel: Interaction coefficient $\  Z \sim \
k^{3/2} \ $ in the nodes of the integer lattice.}
\end{figure}

\section{Examples}

\begin{itemize}

\item{} {\bf Ex.1 } Turbulence of capillary waves (dispersion function $\omega^2=k^3$, three-wave
interactions) was studied in \cite{zak3} in the frame of simplified
dynamical equations for the potential flow of an ideal
incompressible fluid. Coexistence of ZK energy spectra and a set of
discrete waves with constant amplitudes was clearly demonstrated.
The reason why in this case the amplitudes are constant is
following: equation
$$
k_1^{3/2}+k_2^{3/2}=k_3^{3/2}
$$
has no integer solutions \cite{discrete}. It means that there exist
no three-wave resonant interactions among discrete capillary waves,
they take no part in the energy transfer through the wave spectrum
and just keep their energy at the low enough level of nonlinearity.

\item{} {\bf Ex.2 } Similar numerical simulations \cite{zak4} with gravity waves on the surface
of deep ideal incompressible fluid  (dispersion function
$\omega^4=k^2$,  four-wave interactions) show again coexistence of
ZK energy spectra and a set of discrete waves. But in this case
waves amplitudes are not constant any more, discrete waves do
exchange their energy and in fact play major role in the energy
transfer due to the  fact that equation
$$
k_1^{1/2}+k_2^{1/2}=k_3^{1/2}+k_4^{1/2}
$$
has many non-trivial integer solutions \cite{fast} (it is important
in this case that 2-dimensional waves are regraded, i.e.
$k=|\vec{k}|=\sqrt{m^2+n^2}$ with integer $m,n$).

\item{}{\bf Ex.3 } Some recurrent patterns were found in different
atmospheric data sets (rawindsonde time series of zonal wind,
atmospheric angular momentum, atmospheric pressure, etc.) These
large-scale quasi-periodic patterns appear repeatedly at fixed
geographic locations, have periods 10-100 days and are called
intra-seasonal oscillations in the Earth atmosphere. In \cite{intra}
Eq.(\ref{BVE}) (dispersion function $\omega=-2m/n(n+1)$, three-wave
interactions) is studied which is classically regarded as a basic
model of climate variability in the Earth atmosphere. It is shown
that a possible explanation of the intra-seasonal oscillations can
be done in terms of a few specific, resonantly interacting triads of
planetary waves, isolated from the system of all the rest planetary
waves.

{\bf Remark } In contrast to the first two examples, in this case
{\it only discrete layer of turbulence exists}. Indeed, while
$\o=-2m/n(n+1)$ is a rational function, any ratio
$\o(m_i,n_i)/\o(m_j,n_j)$ is a rational number and KAM-theory is not
applicable.

\item {\bf Ex.4 } A very challenging idea indeed is to use the theory of
laminated turbulence to  explain  so-called anomalous energy
transport in tokamaks. Turbulent processes  responsible for these
effects are usually described as H- and L-modes and ELMs (high, low
and edge localized modes consequently). Interpretation of the known
experimental results in terms of  non-resonant (H), resonant (L) and
resonant with small non-zero resonance width (ELM) modes gives
immediately a lot of interesting results. In this case {\it
non-resonant} discrete waves are of major interest because they will
keep their energy as in {\bf Ex.1}, for a substantial period of
time. This approach allows to get two kind of results: 1) to
describe the set of the boundary conditions providing no resonances
at all - say, if the ratio of the sides in the rectangular domain is
2/7,  no exact resonances exist; or 2) to compute explicitly all the
characteristics of the non-resonant waves (wave numbers,
frequencies, etc.) for given boundary conditions. Some preliminary
results are presented in \cite{drift}, in the frame of Hasegawa-Mima
equation  in a plane rectangular domain  with zero boundary
conditions (dispersion function $\omega=1/\sqrt{n^2+m^2}$,
three-wave interactions).

\end{itemize}

It is important to remember that a choice of  initial and/or
boundary conditions for a specific PDE might lead to a substantially
 different form of dispersion function and consequently to the qualitatively
 different behavior of the wave system. For instance, {\bf
 Ex.3}
 and {\bf Ex.4} are described by the same Eq.(\ref{BVE})
regarded on a sphere ({\it rational dispersion function, only
discrete layer of turbulence exists}) and in a rectangular ({\it
irrational dispersion function, both layers exist}) respectively.
For some equations, a special choice of boundary conditions leads to
transcendental dispersion functions.

\section{Open Questions}

We have seen that the main algebraic object of the wave turbulence
theory is the equation \be \label{open}
\o(m_1,n_1)+\o(m_2,n_2)+...+\o(m_s,n_s)=0 \ee where  dispersion
function $\o$ is a solution of a dispersive evolutionary LPDE with
 $ \arrowvert \frac {\partial^2\omega}{\partial k_i
\partial k_j} \arrowvert \neq 0. $  So defined class of dispersion functions includes
rational, irrational or transcendental function, for instance
$$
 \o ( k)=\a k - \b k^3,
\ \
  \o^4 ( k)= \a^2k^2+\b^2, \ \  \o=m/(k+1), \ \  \o=\tanh{\a k}, \ \cdots
$$
where $\a$ and $\b$ are constants and $k=\sqrt{m^2+n^2}$. Continuous
layer of the wave turbulence, that is,  with  $\ m_i, n_i  \in
\Real, \ $ is well studied. On the contrary, there are still a lot
of  unanswered questions concerning the discrete layer of
turbulence,  $m_i, n_i \in \Integer $,
 and we formulated here just a few of them.\\

  Eq.(\ref{open}) can be regarded as {\it a summation rule} for the
rational points of the manifold given by $\o$. These manifolds have
very special structure - namely, they can be transformed into a
one-parametric family of simpler manifolds, namely (\ref{6_irra})
into (\ref{3_ra}). This situation is general enough, the definition
of classes can be generalized for
 a given $c \in \Natural, c \neq 0, 1, -1$ considering
algebraic numbers ${k^{1/c}, k \in \Natural}$ and their unique
representation
$$
    k_c = \g q^{1/c} , \g \in \Integer
$$
where $q$ is a product
$$
    q=p_1^{e_1} p_2^{e_2} ... p_n^{e_n},
$$
while $p_1, ... p_n$ are all different primes and the powers $e_1,
...e_n \in \Natural $ are all smaller than $c$. Then algebraic
numbers with
 the same $q$ form the class $Cl_q$
and the following statement holds:\\

{\it The equation $ a_1 k_1 + a_2 k_2 ...  + a_n k_n = 0, a_i \in
\Integer $ where each $k_i = \g_i q_i^{1/c}$ belongs to some class
$q_i \in q_1, q_2 ... q_l, \quad l<n$ with ...  is equivalent to a
system}
\bea \label{sys2}
\begin{cases}a_{q_1,1} \g_{q_1,1} + a_{q_1,2}
\g_{q_1,2} + ...+ a
_{q_1,n_1} \g_{ q_1,n_1} = 0\\
a_{q_2,1} \g_{q_2,1} + a_{q_2,2} \g_{q_2,2} + ... + a _{q_2,n_2}
\g_{
q_2,n_2} = 0\\
...\\
a_{q_l,1} \g_{q_l,1} + a_{q_l,2} \g_{q_l,2} + ... + a _{q_l,nl} \g_{
q_l,nl} = 0
\end{cases}
\eea

The questions are: what is the geometry underlying this
parametrization? What is known about these sort of manifolds? What
other properties of the resonance manifold are defined by a given
summation rule? What additional information about these manifolds
gives us the fact that they
have many (often infinitely many) integer points?\\

Another group of questions concerns transcendental dispersion
functions. All our examples were constructed for rational and
irrational dispersion functions and the theoretical results were
based on some classical theorems on the linear independence of some
sets of algebraic numbers. In the case of a transcendental
dispersion function like $ \ \o=\tanh{\a k} \ $ similar reasoning
can be carried out using the theorem on the linear independence of
the exponents but it is not done yet. The question about special
functions in this context is completely unexplored though very
important. For instance, a dispersion function for capillary waves
in a circle domain is described by Bessel function. Any results on
their resonant interactions will
shed some light on the nature of Faraday instability.\\

One of the most interesting questions about the resonance manifolds
would be to study  their invariants, i.e. some new function $\
f=f(m,n,\o)$ such that $\ \o_1+\o_2=\o_3 \ $ implies $\ f_1+f_2=f_3.
\ $  An example of this sort of analysis is given in \cite{zak_balk}
for 3-wave interactions of drift waves with $\o=\a x /(1+y^2)$ but
for real-valued wave vectors,  $\ x,y \in \Real. \ $ Existence of
the invariants is important because it is directly connected with
the integrability of corresponding nonlinear PDE. Coming back to the
physical language this means that the wave
system possesses some additional conservation law.\\

A very important task would be to develop  fast algorithms to
compute integer points on the resonance manifolds. The
parametrization property allows to construct specific algorithms for
a given dispersion function as it was done in \cite{fast} for 4-wave
interactions of gravity waves, $\ \o=(m^2+n^2)^{1/4} \ $. The work
on the generic algorithm for a dispersion function $\ \o=\o(k) \ $
($ \ k=\sqrt{m^2+n^2} \ \  \mbox{or} \ \ k=\sqrt{m^2+n^2+l^2} \ $
with integer $\ m,\ n,\ l \ $) is on the way \cite{karkar1} but it
does not cover even simple cases like $\ \o=m/n^2 \ $ not mentioning
transcendental dispersion functions.

\section*{ACKNOWLEDGMENTS}

Author acknowledges support of the Austrian Science Foundation (FWF)
under projects SFB F013/F1304.


\begin{thebibliography}{99}

\bibitem{arnold}  V.I. Arnold.  {\bf Lectures on Partial Differential Equations.}
Springer Series:  Universitext, 157 pp. (2004)


\bibitem{whith1} G. B. Whitham. {\bf Linear and Nonlinear Waves}.  Wiley
Series in Pure and Applied Mathematics, 636 pp. (1999)


\bibitem{ho} M. A. Holmes. {\bf Introduction to Perturbation Methods}.
Springer-Verlag, New York (1995)

\bibitem{lvov} V.E.Zakharov, V.S.L'vov, G. Falkovich. "Kolmogorov Spectra
of Turbulence", Series in Nonlinear Dynamics, Springer (1992)

\bibitem{fil} V.E. Zakharov, N.N. Filonenko. "Weak turbulence of capillary
waves". {\it Zh. Prikl. Mekh. Tekh. Phys.} {\bf 4} (5), pp.62-67
(1967)

\bibitem{KAM}
 A.N. Kolmogorov. "On the conservation of
conditionally periodic motions for a small change in Hamilton´s
function". {\it Dokl. Akad. Nauk SSSR}, {\bf 98}, pp. 527 (1954).
English translation in: {\bf Lecture notes in Physics 93}, Springer
(1979); V.I. Arnold. "Proof of a theorem by A.N. Kolmogorov on the
invariance of quasi-periodic motions under small perturbations of
the Hamiltonian." {\it Russian Math. Surveys}, {\bf 18}, pp.9
(1963);
 J. Moser. "On invariant curves of area preserving mappings of an
annulus." {\it Nachr. Akad. Wiss. Gött., Math. Phys. Kl.}, pp.1-20
(1962)

\bibitem{discrete} E.A. Kartashova. "Partitioning of ensembles of
weakly interacting dispersing waves in resonators into disjoint
classes." {\it Physica D},  {\bf 46} pp.43 (1990); E.A. Kartashova.
"Weakly nonlinear theory of finite-size effects in resonators". {\it
Phys. Rev. Letters},
 {\bf 72}, pp.2013 (1994);
 E.A. Kartashova. "Wave resonances in systems
with discrete spectra". In: V.E.Zakharov (Ed.) {\bf Nonlinear Waves
and Weak Turbulence}, Series: Advances in the Mathematical Sciences,
AMS, pp.95-129 (1998)

\bibitem{tue} Schmidt W.M.  {\bf Diophantine approximations}. Springer Math.
Lecture Notes 785, Berlin, 1980.

\bibitem{JETP} E.A. Kartashova. "A model of laminated wave turbulence".
{\it JETP Letters}, {\bf 83} (7), pp.341 (2006)

\bibitem{zak3} A.N. Pushkarev, V.E. Zakharov. "Turbulence of capillary waves -
theory and numerical simulations." {\it Physica D},  {\bf 135},
pp.98 (2000)

\bibitem{zak4} V.E. Zakharov, A.O.Korotkevich, A.N. Pushkarev,
A.I. Dyachenko. "Mesoscopic wave turbulence",  {\it JETP Letters},
{\bf 82} (8), pp.487 (2005)

\bibitem{fast} E. Kartashova. "Fast Computation Algorithm for Discrete Resonances
among Gravity Waves", {\it JLTP}  to appear. E-print
arXiv.org:nlin/0605067 (2006)

\bibitem{intra} E. Kartashova, V. L'vov. "Large-scale variability in
the Earth atmosphere". E-print arXiv.org:nlin/0606058. Shortened
version submitted to {\it Phys. Rev. Letters} (2006)

\bibitem{drift} E. Kartashova. "Kinetic equation and Clipping - two limits of wave turbulence
theory". E-print arXiv.org:math-ph/0509006 (2005)

\bibitem{zak_balk} A.M. Balk, S.V. Nazarenko, V.E.Zakharov.  "New
invariant for drift turbulence". {\it Phys. Letters A} {\bf 152}
(5-6), pp. 276

\bibitem{karkar1}  Kartashova E., Kartashov A.
{\it Laminated Wave Turbulence: Generic Algorithms I.}
  E-print arXiv.org:math-ph/0609020 . {\bf Submitted} to J. Comp.
  Phys. (2006)
\end{thebibliography}
\end{document}